\documentclass[journal,comsoc]{IEEEtran}

  \usepackage[pdftex]{graphicx}

\usepackage{amsmath}
\usepackage[cmintegrals]{newtxmath}

\usepackage{amsmath}
\usepackage{amsfonts,amssymb}
\usepackage{multirow}
\usepackage{color}
\usepackage{scalerel}
\usepackage{subfig}

\usepackage{tikz}
\usetikzlibrary{tikzmark}
\usetikzlibrary{positioning}

\def\bF{{\mathbb{F}}}

\def\imp{\mathop{\rm imp}}

\begin{document}
\title{Physical Layer Security Protocol for Poisson Channels 
for Passive Man-in-the-middle Attack\thanks{MH was supported in part by JSPS Grant-in-Aid for Scientific Research (A) No.17H01280 and for Scientific Research (B) No.16KT0017, and Kayamori Foundation of Informational Science Advancement.
}}

\author{Masahito~Hayashi 
        and
\'{A}ngeles V\'{a}zquez-Castro, 
\thanks{Masahito Hayashi is with
Graduate School of Mathematics, Nagoya University, Nagoya, Japan,
Shenzhen Institute for Quantum Science and Engineering, Southern University of Science and Technology, Shenzhen, China, 
Center for Quantum Computing, Peng Cheng Laboratory, Shenzhen, China,
and
Centre for Quantum Technologies, National University of Singapore, Singapore.
e-mail: masahito@math.nagoya-u.ac.jp}
\thanks{
\'{A}ngeles Vazquez-Castro
is with the Department of Telecommunications and Systems Engineering, and with The Centre for Space Research (CERES) of Institut d'Estudis Espacials de Catalunya (IEEC-UAB) at 
Autonomous University of Barcelona,
Barcelona, Spain
e-mail: angeles.vazquez@uab.es.}
\thanks{Manuscript submitted 30th June 2018; revised xxx, 2017.}}

\maketitle

\begin{abstract}
In this work, we focus on the classical optical channel having Poissonian statistical behavior and propose a novel secrecy coding-based physical layer protocol. Our protocol is different but complementary to both (computationally secure) quantum immune cryptographic protocols and (information theoretically secure) quantum cryptographic protocols. Specifically, our (information theoretical) secrecy coding protocol secures classical digital information bits at photonic level exploiting the random nature of the Poisson channel.

It is known that secrecy coding techniques for the Poisson channel based on the classical one-way wiretap channel (introduced by Wyner in 1975) ensure secret communication only if 
the mutual information to the eavesdropper 
is smaller than that to the legitimate receiver.
In order to overcome such a strong limitation, we introduce a two-way protocol that always ensures secret communication independently of the conditions of legitimate and eavesdropper channels. We prove this claim showing rigorous comparative derivation and analysis of the information theoretical secrecy capacity of the classical one-way and of the proposed two-way protocols. 
We also show numerical calculations that prove drastic gains and strong practical potential of our proposed two-way protocol to secure information transmission over optical channels. 
\end{abstract}


\section{Introduction}
 \subsection{Rationale and main contributions}
Optical carriers are of increasing relevance for both terrestrial and space applications. For example, 5G communication includes the use of optical wireless channels as part of cellular architecture or for backhauling of wireless networks \cite{5G-1, 5G-2,5G-3}. Optical links are inherently more secure than carriers in the radio frequency range while improvements in throughput are expected to be significant. However, there still exists thread of eavesdropping under realistic channel conditions, which can greatly impair the communication due to the high throughputs.

In this work, we focus on the classical optical channel having Poissonian statistical behavior and propose a secrecy coding protocol that secures classical digital information bits at photonic level exploiting the random nature of the Poisson channel. It is known that secrecy coding techniques for the Poisson channel based on the classical one-way wiretap channel (introduced by Wyner in 1975 \cite{Wyner1975}) ensures secret communication only 
if the mutual information $I_E$ to the eavesdropper
is smaller than that $I_L$ to the legitimate receiver 
with a certain input conditional distribution \cite{Csiszar1978}.
Since this condition means that the positivity of the difference $I_L-I_E$
with a certain input conditional distribution,
we call it the {\it positive information difference condition} for the wiretap channel. 

\begin{figure}[tbh]
\centering
\includegraphics[scale=0.45]{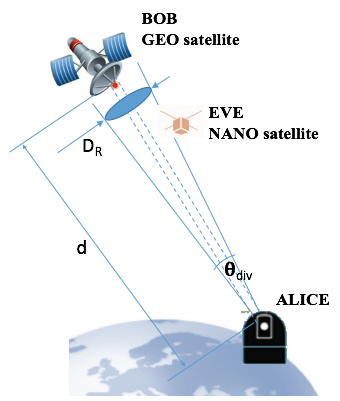}
\protect\caption{Example of ground-to-satellite optical communication scenario where Eve is a nanosatellite not to be detected by intrusion-detection techniques by neither Alice nor Bob.
$\theta_{\textsf{div}}$ is the divergence
angle in radians and ${\textsf d}$ is the distance between transmitter
and receiver.
${\textsf D}_{\textsf R}$ is the aperture diameter.}
\label{fig:OpticalScenario}
\end{figure}

As a typical and illustrative case, we consider a classical optical channel from a ground station (the legitimate transmitter)
to a satellite (the legitimate receiver).
In this case,  as shown in Fig. \ref{fig:OpticalScenario},
there is a possibility that a nano-satellite makes a passive man-in-the-middle attack under the typical beam divergence of an optical transmission beam.
In this case, the mutual information to the eavesdropper cannot be smaller than that to the legitimate receiver.
Hence, the conventional one-way wiretap channel  coding does not work.

In order to overcome this type of man-in-the-middle attack,
we introduce a two-way protocol that always ensures secret communication independently of the conditions of legitimate and eavesdropper Poisson channels. 
Our two-way protocol is given as follows.
In the first round, one legitimate end sends binary data by the on-off keying (OOK) modulation
to the other legitimate end via a Poisson channel.
The later legitimate applies threshold detection to obtain a binary stream of data. Since the eavesdropper makes a passive man-in-the-middle attack, the eavesdropper wiretaps this communication via another Poisson channel, but cannot modify the Poisson channel to the latter legitimate.
Hence, the noises to these two channels are assumed to be independent.
In the second round, the latter legitimate end encodes the message to a binary data,
and sends the former legitimate the exclusive or between the binary data and the binary data obtained in the first round via a public noiseless channel.
The public noiseless channel is assumed to be perfectly eavesdropped by the 
eavesdropper, hence, this model takes the passive man-in-the-middle attack in the second round into account.

In proving this claim and after a brief review of fundamental knowledge of the wiretap channel in Section II, 
we provide a number of novel contributions structured in the different sections as follows. 
In Section III, we present a one-way protocol for the wiretap Poisson channel, with which we will compare our proposed two-way protocol. 
In doing so we first of all formulate signal model as a function of the average number of photons received from the physical channel with 
the OOK modulation and two types of detectors: threshold detector and photon number resolving (PNR) detector.
Although the latter was studied in \cite{Lapidoth,Laourine,ZGX}
and this model with on-off detector (a special case of the former) was studied in \cite{Endo2015} ,
this model with general threshold detector (the general case of the former) is novel.
While 
the channel degradation conditions are known for the latter \cite{Lapidoth,Laourine}, 
we derive the secrecy capacity and channel (stochastic and anti-stochastic) degradation conditions for the former. 
In addition, for the remaining possible conditions of the channel we obtain the secrecy capacity using the general formula assuming channel pre-indexing. 
In Section IV we present a novel two-way protocol for the wiretap Poisson channel (following up the protocol for the wiretap Gaussian channel in \cite{Two,RealisticChannel}, which is based on 
Maurer's idea \cite{Maurer1993}.). 
While the signal model is the same as in the one-way protocol, there are now two rounds of communications. We show that in this case Bob always uses threshold detector, but Eve can use either threshold detector or photon-number-resolving detection. 
For all the cases, the secrecy capacity is shown to be always positive irrespective of the channel conditions of both the legitimate and the eavesdropper channel. Moreover it is also analytically shown that the secrecy rate in the two-way protocol using threshold detector is always not smaller than the secrecy rate in the one-way protocol. 
In Section V, we numerically compare 
the secrecy capacities of our two-way protocols with 
the secrecy capacities of the one-way protocols.
This numerical comparison clarifies how drastically our two-way protocols improve the one-way protocols.
This improvement shows strong practical potential of our proposed two-way protocol to secure information transmission over optical channels.

 \subsection{Novelty compared with existing results}
 The applicability of information theoretical security to enhance secret sharing and key agreement over optical links have been investigated in \cite{Wang2014} while outage probability due to laser-beam divergence and turbulence-induced fading is discussed in \cite{LopezM2014}. 
 The paper \cite{Lapidoth} discussed wiretap Poisson channel only with one-way protocol,
 where they assume the PNS detector.
Recently, the paper \cite{Endo2015} investigated 
the OOK modulation secrecy capacity with the on-off detector
in the context of realistic optical communication.
They obtained upper bounds on the decoding error probability and the leaked information. 
Then, they made useful numerical analysis in a realistic setting 
with a certain quantum regime.
In particular, they clarified the improvement by use of a pre-index channel in their practical setting.
However, all of these works assume the classical one-way wiretap channel introduced by Wyner in 1975 \cite{Wyner1975}, which imposes the well known requirement 
(the positive information difference condition)
of the relation between the two mutual informations to the eavesdropper and the legitimate 
for secrecy provision \cite{Csiszar1978}. 
  
Recently several studies \cite{WGH,FJHWY,ZWPT,TY}  proposed 
two-way protocols to overcome the positive information difference condition. 
However, they require several communication rounds.
The paper \cite{Pierrot2011} addressed two-way protocol with two rounds with noisy feedback.
The papers \cite{Ahlswede,Dai,Gunduz,Li,Wei} discussed the secrecy capacity with secure noiseless feedback.
Since their feedback channel needs to be secure, 
their setting is different from our case.
Using the idea by Maurer \cite{Maurer1993}, our previous papers \cite{Two,RealisticChannel,Correlation} proposed a 
two-way protocol with noiseless public channel for Gaussian BPSK channel.
No paper studied the Poisson-channel case for 
two-way protocol with noiseless public channel
although this model can be applied to the classical optimal communication.
In fact, Gaussian BPSK channel is a symmetric channel, but Poisson channel is not symmetric.
Hence, the protocol for Gaussian BPSK channel \cite{Two,RealisticChannel} cannot be directly applied to our case.  
Using a different protocol, this paper resolves this problem, and 
discusses how a two-way protocol with noiseless public channel
improves the secrecy capacity for Poisson channels.
Our improvement is much larger than that by use of a pre-index channel in one-way protocol.
Further, in our two-way protocol, pre-index does not improve the secrecy capacity.
Since our protocol works without the positive information difference condition for the wiretap channel
and a noiseless public channel allows any information leakage,
it realizes the secure communication even with the man-in-the-middle attack.
Notice that a noiseless public channel can be realized by combination of an error correction
and a conventional wireless channel without any secrecy condition.
Hence, the assumption for a noiseless public channel is more realistic than that for 
secure noiseless feedback.

\section{Wiretap channel}
Before our discussion of optical channel, 
we review fundamental knowledge for wiretap channel.
The channel from the legitimate transmitter (Alice) to the legitimate receiver (Bob) is referred to as the ``main'' channel, and is considered to be a memoryless channel characterized by input alphabet ${\cal X}$, output alphabet ${\cal Y}$, a transition probability $W_{Y |X}$. The other channel from Alice to a passive adversary (Eve) is referred to as the ``eavesdropper's channel'', and consists of another memoryless channel characterized by input alphabet ${\cal X}$,  output alphabet ${\cal Z}$, and transition probability $W_{Z|X}$. 
The pair of channels $(W_{Y|X},W_{Z|X})$ is called wiretap channel.
We assume that both the main and the eavesdropper channel are discrete memoryless channels (DMCs)
of $W_{Y|X}$ and $W_{Z|X}$.
This model supposes that the statistics of both channels are known to all parties, and that authentication is already done\footnote{A small secret key is needed to authenticate the communication. 
It is known that $\log n$ bits of secret are sufficient to authenticate $n$ bits of data~\cite{Wegman1981}}. 
Also, it is assumed that Eve knows the coding scheme used by Alice and Bob. 
For either channel model, the classic wiretap code design goal is the simultaneous provision of reliability and security and requires stochastic encoders. 
Then, 
the limit of secure transmission rate is called the secrecy capacity $C_s$.
To handle $C_s$, we consider an additional system ${\cal V}$ and a joint distribution $p_{VX}$
on ${\cal V}\times {\cal X}$.
Then, we have the joint distributions
$p_{VXY}:=W_{Y|X}p_{XV}$ and $p_{VXZ}:=W_{Z|X}p_{XV}$,
on ${\cal V}\times {\cal X}\times {\cal Y}$ and ${\cal V}\times {\cal X}\times {\cal Z}$, respectively.
Under these distributions, we have
the mutual informations $I(V;Y)$ and $I(V;Z)$. 
Then, we have the formula for the secrecy capacity $C_s$ as \cite{Csiszar1978}
\begin{align}
  \label{eq:secrecy_capacity}
  C_s = \max_{p_{VX}}\left(I(V;Y)-I(V;Z)\right)_+,
\end{align}
where
$(a)_+$ is $a$ for a positive number $a$ and is $0$ for a negative number $a$. 
For example, 
there exists a channel $W_{Z|Y}$ such that
$W_{Z|X}=W_{Z|Y}\circ W_{Y|X}$, 
the wiretap channel $(W_{Y|X},W_{Z|X})$ is called stochastically degraded.
In this case, 
the secrecy capacity $C_s$ 
is written as follows \cite{Wyner1975}.
\begin{align}
  \label{eq:secrecy_capacity2}
  C_s = \max_{p_{X}}\left(I(X;Y)-I(X;Z)\right)_+.
\end{align}
That is, we do not need to consider the additional variable $V$.
Further,
when 
there exists a channel $W_{Y|Z}$ such that
$W_{Y|X}=W_{Y|Z}\circ W_{Z|X}$, 
we call the wiretap channel $(W_{Y|X},W_{Z|X})$ anti-stochastically degraded.
In this case, the secrecy capacity $C_s$ is zero.

When the channels $W_{Y|X}$ and $W_{Z|X}$ satisfy a certain symmetric condition,
the maximum \eqref{eq:secrecy_capacity} is attained with the uniform distribution in 
\eqref{eq:secrecy_capacity2},
and the secrecy capacity $C_s$ can be attained by a combination of randomized error correcting linear codes and 
a random privacy amplification \cite[Section V]{Hayashi2011}.
When the maximum \eqref{eq:secrecy_capacity2}
is attained only by the non-uniform distribution $p_X$,
we need to employ non-linear code to realize a code whose transmission rate attains the capacity.
Generally, it is quite hard to implement such a non-linear code efficiently.

The same problem happen when we employ the formula \eqref{eq:secrecy_capacity}.
To attain the mutual information rate $I(V;Y)-I(V;Z)$,
we apply the above discussion to the channels $W_{Y|V}$ and $W_{Z|V}$. 
When the marginal distribution $P_V$ is uniform, 
the rate can be attained by 
a combination of randomized error correcting linear codes and 
a random privacy amplification. 
Otherwise, we need to employ non-linear code to realize a code.

\section{One-way protocol}
\subsection{Channel model}
In this paper, as a typical channel model of the low-power free space optical (FSO) channel using OOK modulation,
we focus on wiretap Poisson channel.
In OOK, we have the binary input $X$, in which
the ``$X=0$" is represented by zero intensity and ``$X=1$" by positive intensity. 
This model has the legitimate and the eavesdropper, which have 
the average photon counting number 
$\lambda^L_{R|X=x}$ and $\lambda^E_{R|X=x}$
in the following way
when considering a binary modulation input random variable $X$ \cite{Shaik1988}\cite{Book2008};
\begin{align}
\lambda^L_{R|X=x}=\xi x+\zeta, \quad 
\lambda^E_{R|X=x}=\gamma_{gp} \xi x + \gamma_{np} \zeta,\label{EE10}
\end{align}
where now the contributions to the average photon number due to dark and/or light scattering currents are given as $\zeta$ and $\gamma_{np} \zeta$ for the legitimate and the eavesdropper's detectors, respectively. 
Here, the coefficient $\gamma_{gp} $ expresses 
the ratio of the power of pulse received by the eavesdropper to that by 
the legitimate.
When both photo detectors have the same performance, 
$\gamma_{np}$ can be regarded as $1$.
For example, when the eavesdropper makes the man-in-the-middle attack,
the coefficient $\gamma_{gp} $ is greater than $1$.

\begin{table}[htpb]
  \caption{Summary of parameters in wiretap Poisson channel model}
\label{T2}
\begin{center}
  \begin{tabular}{|c|l|} 
\hline
\multirow{2}{*}{$\xi$ }& 
Signal contribution to legitimate's 
\\
&
average photon counting number.\\ 
\hline
\multirow{2}{*}{$\zeta$ }& 
Dark and/or scattered light current contribution
\\
& to legitimate's average photon counting number.
\\
\hline
\multirow{2}{*}{ $\gamma_{np}$ }
&
Eavesdropper's noise coefficient in \\
&
terms of photon counting w.r.t. legitimate's noise.\\
\hline
\multirow{3}{*}{$\gamma_{gp}$ }& 
Eavesdropper's channel attenuation coefficient in \\
&
terms of photon counting w.r.t. legitimate's\\
&
channel attenuation coefficient.
\\
\hline
  \end{tabular}
\end{center}
\end{table}

We now denote the Poisson distribution of the random variable $Y$ 
with average photon counting $\lambda$ by  $\mathcal{P}_{\lambda}(~)$ defined as 
$\mathcal{P}_{\lambda}(y) = \frac{e^{-(\lambda)}(\lambda )^y}{y!}$ for $y \in \mathbb{Z}^+.$ 
The received random variable at the legitimate and eavesdropper's receivers are denoted as $Y$ and $Z$, respectively whose probabilities are controlled by the transition channel probabilities defined as
\begin{align}
W_{Y|X}(y|x) &= \mathcal{P}_{(\xi x + \zeta)}(y)= \mathcal{P}_{\lambda^L_{R|X=x}}(y),
\label{H1}
\\ 
W_{Z|X}(z|x) &= \mathcal{P}_{(\gamma_{gp} \xi x + \gamma_{np} \zeta)}(z) 
= \mathcal{P}_{\lambda^E_{R|X=x}}(z).
\label{H2}
\end{align}
That is, when the legitimate receiver and the eavesdropper use the PNR detector,
they observe $Y$ and $Z$, respectively. 

\subsection{Protocol}
\begin{figure}[tbh]
\centering
\includegraphics[scale=0.45]{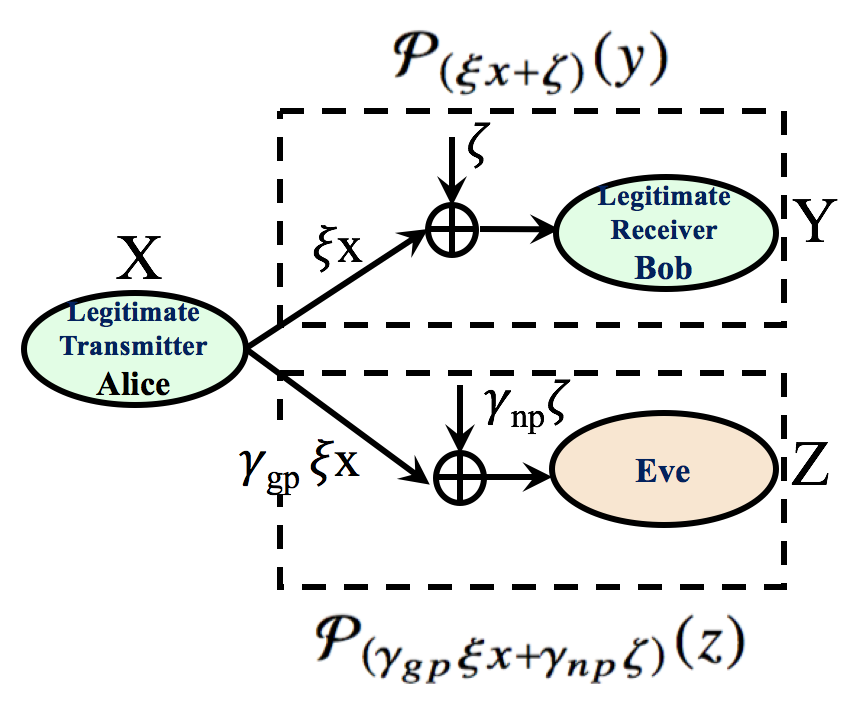}
\protect\caption{Graphical model of the information theoretical one-way optical channel for Poisson distributed received signals according to (\ref{H1}) and  (\ref{H2}). Note that the shown values are the mean values of the Poisson distributions.}
\label{fig:AverageSignals}
\end{figure}

In the one-way model, the secrecy capacity of a memoryless wiretap channel is completely characterized as \eqref{eq:secrecy_capacity}, which is statistically described by the conditional distribution $W_{YZ|X}$. In the following we compute the secrecy capacity of the optical Poisson channel based on the analysis on the conditions of stochastic channel degradation. For the one-way protocol, the legitimate receiver in \eqref{H1} corresponds to Bob, and the eavesdropper in \eqref{H2} corresponds to Eve. A graphical model of the information theoretical one-way optical channel for Poisson distributed received signals according to (\ref{H1}) and  (\ref{H2}) is shown in Fig. \ref{fig:AverageSignals}.
In the following, we assume two different implementations of the photodetector. 

\subsection{Bob and Eve use threshold detectors}\label{S4-B}

When the legitimate receiver and the eavesdropper use threshold detector
with the thresholds $\kappa$ and $\kappa'$, respectively, 
we denote the binary outcomes of the legitimate receiver and the eavesdropper
by $Y'$ and $Z'$, respectively.
Then, 
by using $\tau(\zeta,\kappa):=e^{-\zeta}\sum_{n=0}^{\kappa} \frac{\zeta^n}{n !}$,
the transition channel for the legitimate receiver is given as
\begin{align}
\begin{aligned}
P(Y'=1 | X=0) &= 1 - \tau(\zeta,\kappa),
\\ P(Y'=1 | X=1) &= 1 - \tau(\xi+\zeta,\kappa),\\ 
P(Y'=0| X=0) &= \tau(\zeta,\kappa),\\ 
P(Y'=0 | X=1) & = \tau(\xi+\zeta,\kappa).
\end{aligned}
\end{align}
The transition channel for the eavesdropper's receiver is given as
\begin{align}
\begin{aligned}
P(Z'=1 | X=0) &= 1 - \tau(\gamma_{np} \zeta,\kappa'),\\ 
P(Z'=1 | X=1) &= 1 - \tau(\gamma_{gp} \xi+\gamma_{np} \zeta,\kappa') ,\\ 
P(Z'=0 | X=0) &= \tau(\gamma_{np} \zeta,\kappa'),\\ 
P(Z'=0 | X=1) & = \tau(\gamma_{gp} \xi+\gamma_{np} \zeta,\kappa').
\end{aligned}
\end{align}
In particular, when the thresholds $\kappa$ and $\kappa'$ are $0$,
threshold detector equals on-off detector.

When the input $X$ is subject to the binary distribution $(p,1-p)$
and the threshold of Bob's detector is $\kappa$, 
the mutual information between $X$ and $Y'$
is given as
\begin{align}
\begin{aligned}
I(\xi,\zeta;p)_{\kappa}
:= &
h(p\tau(\zeta,\kappa)+(1-p)\tau(\xi+\zeta,\kappa))\\ 
&-p h(\tau(\zeta,\kappa) )- (1-p) h(\tau(\xi+\zeta,\kappa)).
\end{aligned}
\end{align}
When $\kappa=\kappa'=0$,
$\gamma_{gp}\le 1$, and $\gamma_{np}\ge 1$, 
our wiretap channel $(W_{Y|X},W_{Z|X})$ is stochastically degraded. 
Also, 
when $\kappa=\kappa'$ with general $\kappa$,
$\gamma_{gp}= 1$, and $\gamma_{np}\ge 1$, 
our wiretap channel $(W_{Y|X},W_{Z|X})$ is stochastically degraded. 
(See Appendix \ref{A1}.)\footnote{
When we apply the PNR detector,   
the condition for the stochastically degraded property was discussed in \cite{Lapidoth}\cite[Lemma 1]{Laourine}
However, when we apply the threshold detector,
we have different condition for the stochastically degraded property.
Hence, we need to consider this condition.}. 
When our wiretap channel $(W_{Y|X},W_{Z|X})$ is stochastically degraded,
the secrecy capacity 
$C^{OW}_{\kappa,\kappa'} 
(\gamma_{gp}, \gamma_{np}|\xi,\zeta) $ is calculated as follows. 
\begin{align}
C^{OW}_{\kappa,\kappa'}  (\gamma_{gp}, \gamma_{np}|\xi,\zeta) 
= \max_{p}
(I_{\kappa,\kappa'}  (\gamma_{gp}, \gamma_{np}|\xi,\zeta;p))_+
\label{e45},
\end{align}
where
the secrecy information rate 
$I_{\kappa,\kappa'}  (\gamma_{gp}, \gamma_{np}|\xi,\zeta;p) $ is defined as
\begin{align}
I_{\kappa,\kappa'}  (\gamma_{gp}, \gamma_{np}|\xi,\zeta;p) :=
I(\xi,\zeta;p)_{\kappa}
-I(\gamma_{gp}\xi,\gamma_{np}\zeta;p)_{\kappa'}.
\label{e_secrate}
\end{align}

Conversely, when 
$\kappa=\kappa'=0$,
$\gamma_{gp}\ge 1$, and $\gamma_{np}\le 1$, 
our wiretap channel $(W_{Y|X},W_{Z|X})$ is anti-stochastically degraded.
Also,
when 
$\kappa=\kappa'$,
$\gamma_{gp}= 1$, and $\gamma_{np}\le 1$, 
our wiretap channel $(W_{Y|X},W_{Z|X})$ is anti-stochastically degraded.
Hence, the secrecy capacity $C^{OW}_{\kappa,\kappa'}  (\gamma_{gp}, \gamma_{np}|\xi,\zeta) $ is zero.

When 
our wiretap channel $(W_{Y|X},W_{Z|X})$ is not 
stochastically degraded nor anti-stochastically degraded,
we need to apply the general formula \eqref{eq:secrecy_capacity} instead of \eqref{eq:secrecy_capacity2}. 
That is, there is a possibility that use of pre-index channel
enhances secrecy information rate.
For this case,
Endo et al. \cite{Endo2015} restricted the number $|{\cal V}|$ of elements of ${\cal V}$ to two 
in this setting when $\kappa=\kappa'=0$, i.e., the case with the on-off detector.
As shown in Appendix \ref{A2}, 
when the number $|{\cal X}|$ of elements of ${\cal X}$ is $\ell$,
the maximum in \eqref{eq:secrecy_capacity}
can be achieved with the case of $|{\cal V}|=\ell$.
Given two probabilities $p_0,p_1,p \in [0,1]$, we choose 
the distribution $P_V$ 
and
the transition matrix $P_{X|V}$ to describe a pre-index channel
as
\begin{align}
\begin{aligned}
&P_{X|V}(0|i)=p_i,\quad
P_{X|V}(1|i)=1-p_i \\
&P_V(0)=p, \quad
P_V(1)=1-p.
\end{aligned}
\end{align}
Then, the mutual information is given as
\begin{align}
\begin{aligned}
&I(\xi,\zeta;p_0,p_1,p)_{\kappa} \\
:= &
h(p \tau(\xi,\zeta,\kappa,p_0)+(1-p)\tau(\xi,\zeta,\kappa,p_1))\\
&-p h(\tau(\xi,\zeta,\kappa,p_0) )- (1-p) h(\tau(\xi,\zeta,\kappa,p_1)),
\end{aligned}
\end{align}
where $\tau(\xi,\zeta,\kappa,p_i):= 
p_i \tau(\zeta,\kappa)+(1-p_i)\tau(\xi+\zeta,\kappa)$.
Hence, the secrecy capacity $C^{OW}_{\kappa,\kappa'} 
(\gamma_{gp}, \gamma_{np}|\xi,\zeta) $ can be calculated as follows. 
\begin{align}
C^{OW}_{\kappa,\kappa'}  (\gamma_{gp}, \gamma_{np}|\xi,\zeta) 
= \max_{p}
(I_{\kappa,\kappa'}^{\imp}  (\gamma_{gp}, \gamma_{np}|\xi,\zeta;p) 
)_+,
\label{e45B}
\end{align}
where
the improved secrecy information rate by pre-index channel
$I_{\kappa,\kappa'}^{\imp}  (\gamma_{gp}, \gamma_{np}|\xi,\zeta;p) $ is defined as
\begin{align}
\begin{aligned}
& I_{\kappa,\kappa'}^{\imp}  (\gamma_{gp}, \gamma_{np}|\xi,\zeta;p) 
\\ 
:=&
\max_{p_0,p_1}
I(\xi,\zeta;p_0,p_1,p)_{\kappa}
-I(\gamma_{gp}\xi,\gamma_{np}\zeta;p_0,p_1,p)_{\kappa'}.
\end{aligned}
\label{e_imp_secrate}
\end{align}

\subsection{Bob and Eve use PNR detectors}
Next, we assume that Bob and Eve use a receiver with the PNR detector. From a design point of view this is a more expensive alternative if compared with the hard-detection receiver.
In this case, the detectors directly measure $Y$ and $Z$ whose distributions are given in 
\eqref{H1} and \eqref{H2}.
Since 
the mutual information between $X$ and $Y$
is $H(Y)-H(Y|X)$,
by cancelling the factor $\log n !$,
it is calculated as 
the following quantity.
\begin{align}
\begin{aligned}
&I(\xi,\zeta;p)_{\rm PNR}\\ 
:= &
h(\xi,\zeta;p)_{\rm PNR}
-ph(\xi,\zeta;1)_{\rm PNR}
-(1-p)h(\xi,\zeta;0)_{\rm PNR},
\end{aligned}
\label{KK1}
\end{align}
where
\begin{align*}
\begin{aligned}
h(\xi,\zeta;p)_{\rm PNR}
:= &
-\sum_{n=0}^{\infty}
\frac{p \lambda_{0}^ne^{- \lambda_{0}}+(1-p)\lambda_{1}^ne^{- \lambda_{1}}}{n !}
\\&\hspace{5ex}
\cdot \log (p{\lambda_{0}^n e^{- \lambda_{0}}} 
+ (1-p){\lambda_{1}^n e^{- \lambda_{1}}}),
\end{aligned}
\end{align*}
where $\lambda_0:= \zeta $ and $\lambda_1:=\xi+ \zeta$.

When $\gamma_{gp} \le  \gamma_{np}$ and $\gamma_{gp} \le 1$,
our wiretap channel $(W_{Y|X},W_{Z|X})$ is stochastically degraded
\cite{Lapidoth}\cite[Lemma 1]{Laourine}.
Then, 
the secrecy capacity $C^{OW}_{\rm PNR} (\gamma_{gp}, \gamma_{np}|\xi,\zeta) $ 
 is given as follows.
\begin{align}
C^{OW}_{\rm PNR} (\gamma_{gp}, \gamma_{np}|\xi,\zeta)
=& \max_{p}
(I_{\rm PNR} (\gamma_{gp}, \gamma_{np}|\xi,\zeta;p) )_+,
\label{MHP}\\
I_{\rm PNR} (\gamma_{gp}, \gamma_{np}|\xi,\zeta;p) 
:= &
I(\xi,\zeta;p)_{\rm PNR}
-I(\gamma_{gp}\xi,\gamma_{np}\zeta;p)_{\rm PNR}.\nonumber 
\end{align}

Conversely, 
when $\gamma_{gp} \ge  \gamma_{np}$ and $\gamma_{gp} \ge 1$,
$W_{Y|X}$ can be considered as stochastically degraded of the channel 
$W_{Z|X}$.
Hence, the secrecy capacity $C^{OW}_{\rm PNR} (\gamma_{gp}, \gamma_{np}|\xi,\zeta) $ is zero.
 
When  our wiretap channel $(W_{Y|X},W_{Z|X})$ is not stochastically degraded nor anti-stochastically degraded,
we need to apply the general formula \eqref{eq:secrecy_capacity} instead of \eqref{eq:secrecy_capacity2}. 
For completion, we have derived the corresponding expressions of secrecy information rate and capacity as follows.
For this case,
given two probabilities $p_0,p_1,p \in [0,1]$, 
the mutual information is given as
\begin{align}
I(\xi,\zeta;p_0,p_1,p)_{\rm PNR} 
:= &
h(\xi,\zeta; p_0p+p_1p)_{\rm PNR}
-ph(\xi,\zeta;p_0)_{\rm PNR}\nonumber \\
&-(1-p)h(\xi,\zeta;p_1)_{\rm PNR},
\end{align}
Hence, the secrecy capacity $C^{OW}_{\kappa,\kappa'} 
(\gamma_{gp}, \gamma_{np}|\xi,\zeta) $ can be calculated as follows. 
\begin{align}
C^{OW}_{\rm PNR} (\gamma_{gp}, \gamma_{np}|\xi,\zeta) 
= &\max_{p}
(I^{\imp}_{\rm PNR} (\gamma_{gp}, \gamma_{np}|\xi,\zeta;p))_+,
\label{e45B2}
\end{align}
where
\begin{align*}
&I^{\imp}_{\rm PNR} (\gamma_{gp}, \gamma_{np}|\xi,\zeta;p) \nonumber\\
:= &\max_{p_0,p_1}
I(\xi,\zeta;p_0,p_1,p)_{\rm PNR}
-I(\gamma_{gp}\xi,\gamma_{np}\zeta;p_0,p_1,p)_{\rm PNR}.
\end{align*}
In fact, when 
the legitimate and the eavesdropper's detectors have the same
the dark and/or light scattering currents, i.e., $\gamma_{np} =1$,
the paper \cite{WynerPoisson, Laourine, ZGX} derived the secrecy capacity as
\begin{align}
&C^{OW}_{\rm PNR} (\gamma_{gp}, 1|\xi,\zeta) 
= 
\max_{p} (g(\xi,1-p)-g(\gamma_{gp} \xi,1-p))_+,
\label{e45A} 
\end{align}
where
\begin{align}
g(\xi,q) := &
q(\xi+\zeta)\log  (\xi+\zeta)
-(q \xi+\zeta)\log  (q \xi+\zeta)\nonumber \\
&+(1-q) \zeta \log \zeta.
\end{align}

\section{Two-way protocol}
\subsection{Protocol}
Similar to \cite{Maurer1993}, this section considers two-way secure communication protocol.
Our two-way protocol for the optical Poisson channel and OOK modulation can be described as follows. 
In an initial step,
Bob generates the binary variable $B \in \bF_2$ subject to the binary distribution $(p,1-p)$, and
sends it to Alice, where $0$ is generated with probability $p$.
Hence,
the legitimate receiver in \eqref{H1} corresponds to Alice,
and 
the eavesdropper in \eqref{H2} corresponds to Eve.
That is, Alice and Eve obtain the variables $Y'$ and $Z'$, where
$Y'$ and $Z'$ are subject to the Poisson distribution $\mathcal{P}_{\lambda^L_{R|B=b}}$ and 
$\mathcal{P}_{\lambda^E_{R|B=b}}$ with
\begin{align}
\lambda^L_{R|B=b}
 = \xi b +\zeta; \quad
\lambda^E_{R|B=b}
=\gamma_{gp} \xi b +\gamma_{np}\zeta.
\label{eq:GaussChannelMMM}
\end{align}
Here, when $B$ is fixed to a certain value, $Y'$ and $Z'$ are assumed to be independent.

\begin{figure}[tbh]
\centering
\includegraphics[scale=0.45]{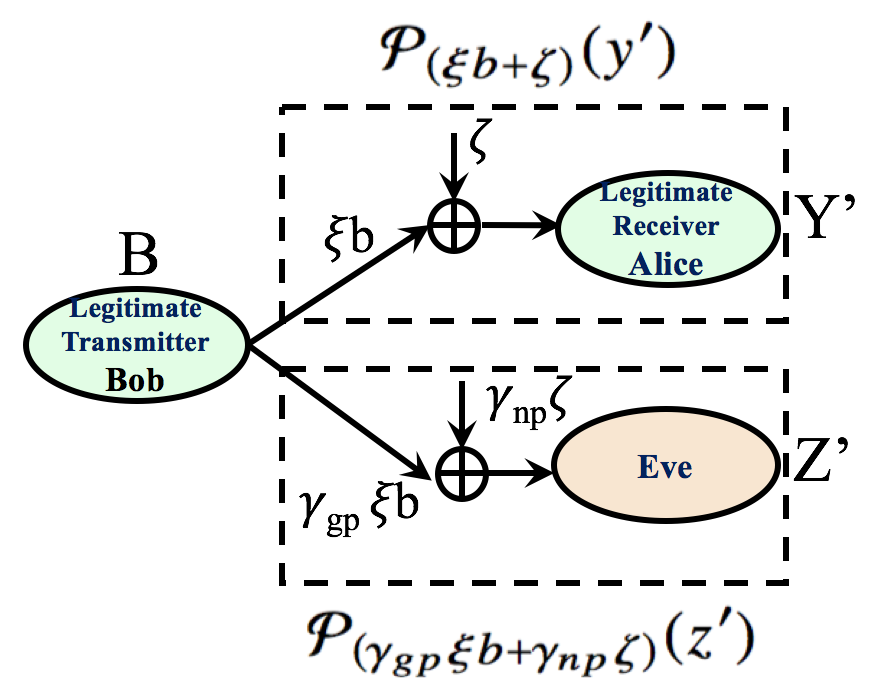}
\protect\caption{Graphical model of the information theoretical optical two-way channel for Poisson distributed received signals according to (\ref{H1}) and  (\ref{H2}). Note that the shown values are the mean values of the Poisson distributions.}
\label{fig:Graphical_twoway}
\end{figure}

Then, applying hard decision decoding to $Y'$
based on threshold detector with threshold $\kappa$, 
Alice obtains the binary variable $A$. 
In the second round, 
Alice prepares another binary variable $X$, and sends $X':= X\oplus A$ to Bob via a public channel from Alice to Bob.
When $X$ is regarded as the channel input information, the legitimate receiver's output is $B$ and $X'$ while the eavesdropper's output is $Z'$ and $X'$. A graphical model of the overall process along with the generated random variables is shown in Fig. \ref{fig:Graphical_twoway}.

\subsection{Secrecy capacity when Eve uses threshold detector}
In the case Eve uses threshold detector with threshold $\kappa'$ 
when receiving $Z'$, 
Eve obtains the binary variable $E$.
Then, we obtain the Markov chain $A-B-E$.
When Alice's input is $X$,
Bob's output is $(B,X')$ and 
Eve's output is $(E,X')$. 
Then, we have the wiretap channel 
$(W_{(B,X'|X)},W_{(E,X'|X)})$.
As discussed in \cite[Eq. (22)]{Hayashi2011},
this channel can be regarded as a symmetric channel.
Hence, the secrecy capacity is achievable by a combination of randomized error correcting linear codes 
in a vector space over the finite field $\bF_2$ and 
a random privacy amplification \cite[Section V]{Hayashi2011}
even when the binary distribution $(p,1-p)$ of the variable $B$ is not uniform.
This is a big advantage over the one-way protocol
because the channel of our one-way protocol is not symmetric, which implies that
a non-linear code is required to attain the capacity in general.
Then, the secrecy capacity equals the symmetric rate and 
is calculated to $I(A;B)-I(A;E)$ \cite[Final equation in Section VI]{Hayashi2011}, which equals to the conditional mutual information 
$I(A;B|E)$
due to the Markov chain $A-B-E$ and has the following expression
\begin{align}
&I(A;B|E)\nonumber \\
=& \sum_{e=0}^1 P_E(e) 
\Big(
h( P_{B|E}(0|e) P_{A|B}(0|0)+ P_{B|E}(1|e) P_{A|B}(0|1)) \nonumber \\
&-P_{B|E}(0|e) h(  P_{A|B}(0|0)) - P_{B|E}(1|e) h(P_{A|B}(0|1))
\Big).
\end{align}
Choosing 
$\bar{q}_0:= p \tau(\lambda^E_{R|B=0},\kappa')
+(1-p) \tau(\lambda^E_{R|B=1},\kappa')$, 
$\bar{q}_1:= 1-\bar{q}_0$,
$\bar{p}_0:=p \tau(\lambda^E_{R|B=0},\kappa')/\bar{q}_0$, and
$\bar{p}_1:=p (1-\tau(\lambda^E_{R|B=0},\kappa'))/\bar{q}_1$,
we have 
$P_E(e) =\bar{q}_e$
and
$P_{B|E}(0|e)=\bar{p}_e$.
Hence, the conditional mutual information is calculated as
\begin{align}
C^{TW}_{\kappa,\kappa',p} (\gamma_{gp}, \gamma_{np}|\xi,\zeta) 
:= \sum_{i=0}^1
\bar{q}_i I(\xi,\zeta;\bar{p}_i)_{\kappa}. 
\end{align}
In this setting, we can optimize the parameter $p$.
Hence, the optimum two-way secrecy capacity is 
\begin{align}
C^{TW}_{\kappa,\kappa'} (\gamma_{gp}, \gamma_{np}|\xi,\zeta)
:=
\max_p
C^{TW}_{\kappa,\kappa',p} (\gamma_{gp}, \gamma_{np}|\xi,\zeta).
\label{e45B3}
\end{align}
Since the analysis in \cite{Hayashi2011} guarantees the strong security,
our protocol realizes the strong security with the above rate.

\begin{figure}[tbh]
\centering
\begin{tikzpicture}[transform shape, node distance = 5cm, auto, every node/.style={outer sep=0}]
    \node [circle,draw] (1) {$B$};
    \node [left =3em of 1,circle,draw](2) {$A$};
    \node [right =3em of 1,circle,draw](3) {$E$};
    \node [above =2em of 1,circle,draw](4) {$U$};
   \draw (2) -> (1) -> (3);
   \draw (1) -> (4);
\end{tikzpicture}\\
\protect\caption{Markov chain.}
\label{Markov}
\end{figure}
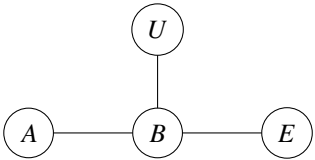

Remember that the parameter $p$ expresses the distribution of the variable $B$. 
One might consider that use of another variable $U$ connected to $B$
as Fig. \ref{Markov}
improves the secrecy rate in a way similar to Section \ref{S4-B}
even in this case.
However, the Markov chain guarantees 
the relation $I(A;B|E) \ge I(A;U|E) $,
which shows that any pre-index channel does not improve 
the performance of the two-way method.

Also, since $I(A;U|E)=I(A;U)-I(A;E)\ge I(U;A)-I(U;E)$, we have
\begin{align}
I(A;B|E) \ge I(U;A)-I(U;E).
\end{align}
Taking the maximum with respect to $p,p_0,p_1$, we have
\begin{align}
C^{TW}_{\kappa,\kappa'} (\gamma_{gp}, \gamma_{np}|\xi,\zeta)
\ge
C^{OW}_{\kappa,\kappa'} (\gamma_{gp}, \gamma_{np}|\xi,\zeta).
\label{MKY}
\end{align}
That is, the two-way method improves the one-way method
even when a pre-index channel is applied in the one-way method.

\subsection{Secrecy capacity when Eve uses PNR decision}
Next, we consider the case when Eve uses the PNR detector.
Then, we obtain the Markov chain $A-B-Z'$.
When Alice's input is $X$,
Bob's output is $(B,X')$ and 
Eve's output is $(Z',X')$. 
Then, we have the wiretap channel 
$(W_{(B,X'|X)},W_{(Z',X'|X)})$, which is also a stochastically degraded channel.  
Similar to the above case,
the secrecy capacity is achievable by the same way 
and is calculated to $I(A;B)-I(A;Z')$ \cite[Final equation in Section VI]{Hayashi2011}, 
which equals to $I(A;B|Z')$
due to the Markov chain $A-B-Z'$.
This value is calculated as
\begin{align}
&C^{TW}_{\kappa, {\rm PNR},p} (\gamma_{gp}, \gamma_{np}|\xi,\zeta)
:= \sum_{n=0}^\infty
q_n I(\xi,\zeta;p_n)_{\kappa},
\end{align}
where 
$q_n:= p \frac{{\lambda^E_{R|B=0}}^n}{n!}e^{-\lambda^E_{R|B=0}}
+(1-p) \frac{{\lambda^E_{R|B=1}}^n}
{n!} e^{-\lambda^E_{R|B=1}} $, and
$p_n:=p \frac{{\lambda^E_{R|B=0}}^n}{n!}e^{-\lambda^E_{R|B=0}}/q_n$.
In this setting, we can optimize the parameter $p$.
Hence, the optimum two-way secrecy capacity is 
\begin{align}
C^{TW}_{\kappa, {\rm PNR}} (\gamma_{gp}, \gamma_{np}|\xi,\zeta)
:=
\max_p
C^{TW}_{\kappa, {\rm PNR},p} (\gamma_{gp}, \gamma_{np}|\xi,\zeta).
\label{e45B4}
\end{align}
Since $E$ is obtained by application of the threshold detector to the variable $Z'$, 
we have $I(A;Z') \ge I(A;E)$, which implies $ I(A;B) -I(A;Z') \le I(A;B) -I(A;E)$.
Hence, we have
\begin{align}
C^{TW}_{\kappa, {\rm PNR}} (\gamma_{gp}, \gamma_{np}|\xi,\zeta)
\le
C^{TW}_{\kappa, \kappa'} (\gamma_{gp}, \gamma_{np}|\xi,\zeta).
\end{align}
Therefore, 
$C^{TW}_{\kappa, \kappa'} (\gamma_{gp}, \gamma_{np}|\xi,\zeta)$
is the largest among three quantities
$C^{TW}_{\kappa, {\rm PNR}} (\gamma_{gp}, \gamma_{np}|\xi,\zeta)$,
$C^{TW}_{\kappa, \kappa'} (\gamma_{gp}, \gamma_{np}|\xi,\zeta)$, and 
$C^{OW}_{\kappa, \kappa'} (\gamma_{gp}, \gamma_{np}|\xi,\zeta)$.

\section{Comparison}
\subsection{Comparison based on channel models \eqref{H1}, \eqref{H2}}
Now, using \eqref{e45B}, \eqref{e45B2}, \eqref{e45B3}, and \eqref{e45B4}, all of these
based on the same wiretap channel model \eqref{H1}, \eqref{H2},
we compare the one-way secure communication with on-off detection and PNR detection 
and two-way secure communication with on-off detection and PNR detection 
considering the asymmetric direction of transmission
because the performance of the two-way secure communication
depends on the property of the initial optical Poisson channel.
As a typical case of threshold detector, we focus on on-off detection. 

It is clear that the legitimate and the eavesdropper choose the minimum value for their dark count rate among their choices because smaller dark count rate improves their information gain.
That is, it is natural to fix $\gamma_{np}$ and $\zeta$ to be certain values.
The parameter $\gamma_{gp}$ depends on the location of the eavesdropper 
when the location of the legitimate signal receiver is fixed. 
Hence, it cannot be controlled by the legitimate.
Since the eavesdropper may move, 
we consider that the parameter $\gamma_{gp}$ changes among a certain range.
Since the parameter $\xi$ depends on the power of the signal,
it is natural that the legitimate signal transmitter can control 
$\xi$ among a certain range.
In this case, even when other parameters $\gamma_{np}$, $\gamma_{gp}$ and $\zeta$
are fixed, it is not so clear what amount is better for $\xi$.

For this comparison, we set $\gamma_{np}=1 $, which means that 
Eve's detector has the same dark count rate as the detector of the legitimate user.
Then, if and only if $\gamma_{gp}\le 1 $,
the channel is stochastically degraded.
Otherwise, 
the channel is anti-stochastically degraded.
Hence, 
regardless of the value of $\gamma_{gp}$,
the one-way secrecy capacities
$C^{OW}_{0,0} (\gamma_{gp}, 1|\xi,\zeta) $ and
$C^{OW}_{\rm PNR} (\gamma_{gp}, 1|\xi,\zeta) $ are calculated 
by \eqref{e45} and \eqref{MHP}, respectively.
That is, we do not need to optimize the additional parameter $p_0$ and $p_1$ related to the pre-index channel
while such a type of optimization is not needed in the two-way protocol.
Hence, if and only if $\gamma_{np}\le 1 $,
both one-way secrecy capacities have positive values.
In contrast, the optimum of the two-way secrecy capacities
$C^{TW}_{0, 0} (\gamma_{gp}, 1|\xi,\zeta) $ and
$C^{TW}_{0, {\rm PNR}} (\gamma_{gp}, 1|\xi,\zeta) $ 
are always positive. 

\begin{figure}[tbh]
\centering
\includegraphics[scale=0.42]{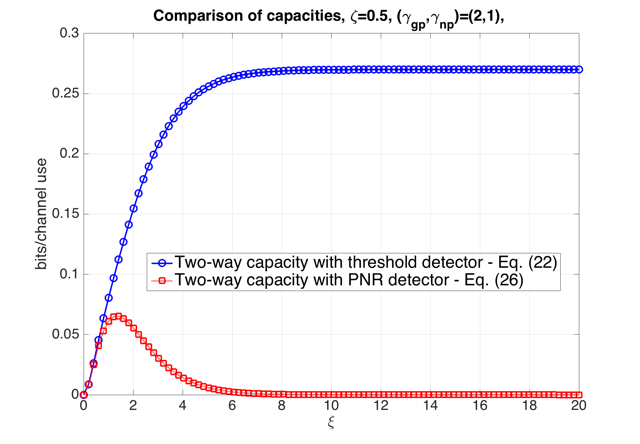}
\protect\caption{
Two-way secrecy capacities 
when $\kappa=0$, $\gamma_{np}=1$, 
$\gamma_{gp}=2$, and $\zeta=0.5$.
The horizontal axis expresses $\xi$ and 
the vertical axis expresses the secrecy capacities.
In this case, one-way secrecy capacities are zero.
Two-way secrecy capacity $C^{TW}_{0, {\rm PNR}} (2, 1|\xi,0.5) $ with PNS detector
has the maximum value when $\xi= 1.35$
while
two-way secrecy capacity $C^{TW}_{0, 0} (2, 1|\xi,0.5) $ with on-off detector
is monotonically increasing for $\xi$.}
\label{Pasted-1}
\end{figure}

In the following, we assume cheaper detector with dark count rate $\zeta=0.5$ unlike the reference \cite{Endo2015}.
When the eavesdropper makes the man-in-the-middle attack,
the eavesdropper receives a stronger pulse than the legitimate.
As a typical case, we assume that the strength of 
the pulse received by the eavesdropper is twice of that by the legitimate, i.e.,
$\gamma_{gp}=2$.
Then, we numerically calculate
the two-way secrecy capacity $C^{TW}_{0, {\rm PNR}} (2, 1|\xi,\zeta) $ 
with PNR detection by varying $\xi$.
The optimum parameter for $\xi$ is $1.35$ as Fig. \ref{Pasted-1}.
In this case, 
when $\xi$ goes to infinity, 
we have 
\begin{align*}
&P_\xi(A=0|B=0)=
P_\xi(E=0|B=0)=e^{-0.5},\nonumber \\
&\lim_{\xi\to \infty}P_\xi(A=0|B=1)=\lim_{\xi\to \infty}P_\xi(E=0|B=1)
=0.
\end{align*}
Hence, the capacity $C^{TW}_{0, 0} (2, 1|\xi,0.5) $ converges to the above case.
However, when Eve uses the PNR detector,
Eve can perfectly distinguish $B=0$ or $1$ under the limit $ \xi \to \infty$. 
In this case, $I(A;E)$ converges to $I(A;B)$ under this limitation, which implies
the capacity $C^{TW}_{0, {\rm PNR}} (2, 1|\xi,0.5) $ approaches to zero.

\begin{figure}[tbh]
\centering
\includegraphics[scale=0.4]{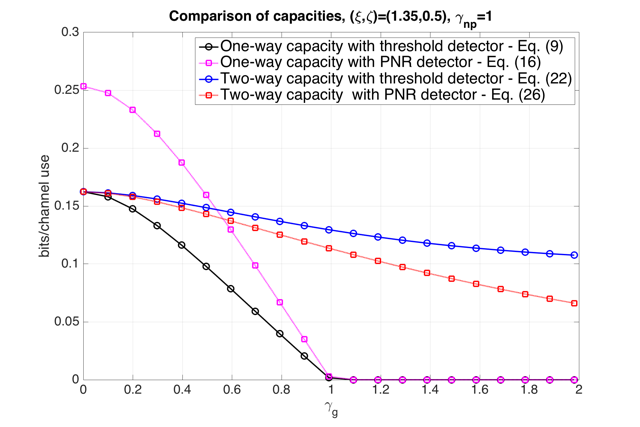}
\protect\caption{
One-way and two-way secrecy capacities 
when $\kappa=\kappa'=0$,
$\gamma_{np}=1$, $\zeta=0.5$, and $\xi=1.35$.
The horizontal axis expresses $\gamma_{gp}$ and 
the vertical axis expresses the secrecy capacities.}
\label{Pasted-2}
\end{figure}

Then,
we set $\xi$ to be $1.35$ while keeping $\gamma_{np}=1 $ and $\zeta=0.5$.
Then, we numerically compare \eqref{e45B}, \eqref{e45B2}, \eqref{e45B3}, and \eqref{e45B4}, 
as Fig. \ref{Pasted-2}.
This comparison shows that 
the two-way secrecy capacity
$C^{TW}_{0, {\rm PNR}} (\gamma_{gp}, 1|1.35,0.5) $ 
is larger than 
the one-way secrecy capacity
$C^{OW}_{0, {\rm PNR}} (\gamma_{gp}, 1|1.35,0.5) $ 
when $\gamma_{gp}$ is larger than 0.5626
while
the two-way secrecy capacity
$C^{TW}_{0, 0} (\gamma_{gp}, 1|1.35,0.5) $ 
is always larger than 
the one-way secrecy capacity
$C^{OW}_{0, 0} (\gamma_{gp}, 1|1.35,0.5) $. 
Further, the difference between 
$C^{TW}_{0, {\rm PNR}} (\gamma_{gp}, 1|1.35,0.5) $ 
and $C^{TW}_{0, 0} (\gamma_{gp}, 1|1.35,0.5) $ 
is not so large.
That is, even when the eavesdropper uses the PNR detector,
the power of the eavesdropper does not increase so much.

\begin{figure}[tbh]
\centering
\includegraphics[scale=0.4]{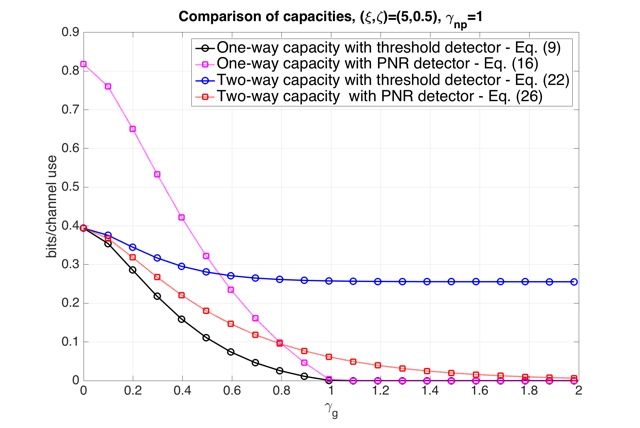}
\protect\caption{
One-way and two-way secrecy capacities 
when $\kappa=\kappa'=0$,
$\gamma_{np}=1$, $\zeta=0.5$, and $\xi=5$.
The horizontal axis expresses $\gamma_{gp}$ and 
the vertical axis expresses the secrecy capacities.}
\label{fig:B}
\end{figure}

Next, since the parameter $\xi$ is under the legitimate control and can actually have larger values, 
we now set $\xi$ to be $5$ while keeping $\gamma_{np}=1 $ and $\zeta=0.5$.
Fig. \ref{fig:B} shows the numerical comparison of these four types of secrecy capacities.
This comparison shows that 
the two-way secrecy capacity
$C^{TW}_{0, {\rm PNR}} (\gamma_{gp}, 1|5,0.5) $ 
is larger than the one-way secrecy capacity
$C^{OW}_{0, {\rm PNR}} (\gamma_{gp}, 1|5,0.5) $ 
when $\gamma_{gp}$ is larger than 0.8014
while
the two-way secrecy capacity
$C^{TW}_{0, 0} (\gamma_{gp}, 1|5,0.5) $ 
is always larger than 
the one-way secrecy capacity
$C^{OW}_{0, 0} (\gamma_{gp}, 1|5,0.5) $. 
Further, the difference between 
$C^{TW}_{0, {\rm PNR}} (\gamma_{gp}, 1|5,0.5) $ 
and $C^{TW}_{0, 0} (\gamma_{gp}, 1|5,0.5) $ is quite large.

From the numerical results shown in Figs. \ref{Pasted-2} and \ref{fig:B}, we observe that from the capacity point of view, when the detector of the eavesdropper is the PNR detector, the choice between using one-way or two-way schemes depends on the value of $\gamma_{gp}$, i.e. on the assumption over the eavesdropper's location. On the other hand, the two-way secrecy capacity is always 
better than the one-way secrecy capacity even with a pre-index channel
when the detector of the eavesdropper is the on-off detector.

In the two-way scheme, 
we fix Bob's detector to be the threshold detector.
Hence, 
the capacity with Eve's threshold detector
is larger than the capacity with Eve's PNR detector.
In the one-way scheme, 
we do not fixed Bob's detector, which is chosen to be the same type of 
Eve's detector.
That is, when Eve uses PNR detector, Bob is assumed to use PNR detector.
Hence,  there is no definitive relation between the capacities with 
the PNR detector and the threshold detector.
The obtained numerical result shows that 
the capacity with the PNR detector 
is larger than the capacity with the threshold detector.
Indeed,
the mutual information 
between Alice and Bob
in the  one-way case with the PNR detector 
(which is denoted by $I_{{\rm PNR}}(A;B)$)
is larger than that in other cases,
(which is denoted by $I_{{\rm THR}}(A;B)$).
Hence, when $\gamma_{gp}$ is very small,
the information leaked to Eve is very small
so that the mutual information between Alice and Bob is dominant.
In this case, the one-way capacity with the PNR detector is close to 
$I_{{\rm PNR}}(A;B)$
and other capacities are close to the same value $I_{{\rm THR}}(A;B)$.
Hence, the one-way capacity with the PNR detector 
is larger than other capacities.

\begin{figure}[tbh]
\centering
\subfloat[$\xi=1.35$]{
\includegraphics[scale=0.35]{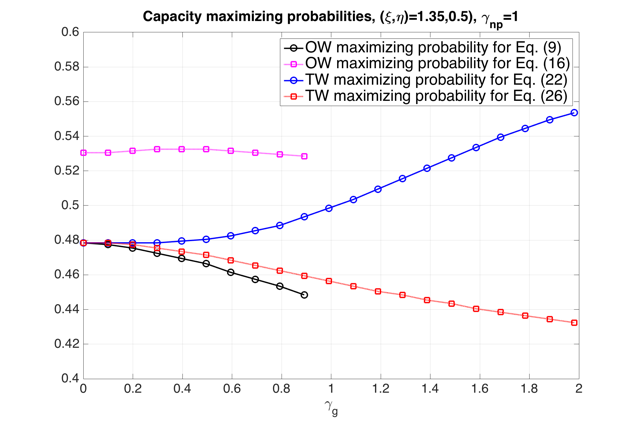}
\label{fig:label-A}}
\\
\subfloat[$\xi=5$]{
\includegraphics[scale=0.35]{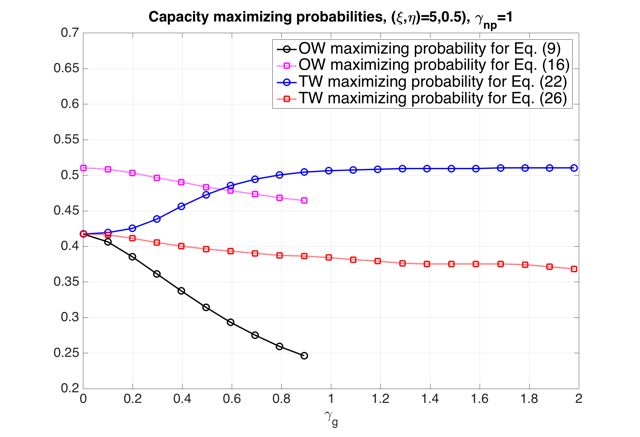}
\label{fig:label-B}}
\caption{Optimal distribution $p$ for one-way and two-way secrecy channel when $\kappa=\kappa'=0$,
$\gamma_{np}=1$, $\zeta=0.5$ and $\xi=1.35$ or $5$.
The horizontal axis expresses $\gamma_{gp}$ and 
the vertical axis expresses the optimal distribution $p$.}
\label{fig:C}
\end{figure}

It is also relevant to analyze the probabilities that maximize the four capacities. We observe in Fig. \ref{fig:C} (a) that when $\xi=1.35$, $\zeta=0.5$ and $\gamma_{np}=1$, all the optimal distributions $(p,1-p)$ 
to achieve the secrecy capacities
$C^{TW}_{0, {\rm PNR}} (\gamma_{gp}, 1|1.35,0.5) $, 
$C^{OW}_{\rm PNR} (\gamma_{gp}, 1|1.35,0.5) $, 
$C^{TW}_{0, 0} (\gamma_{gp}, 1|1.35,0.5) $, 
and
$C^{OW}_{0, 0} (\gamma_{gp}, 1|1.35,0.5) $
are close to the uniform distribution.
In contrast, when $\xi=1.35$ is changed to $5$,
Fig. \ref{fig:C} (b)
shows the optimal distribution $(p,1-p)$ 
to achieve the secrecy capacities
$C^{TW}_{0, {\rm PNR}} (\gamma_{gp}, 1|5,0.5) $, 
$C^{OW}_{\rm PNR} (\gamma_{gp}, 1|5,0.5) $, 
$C^{TW}_{0, 0} (\gamma_{gp}, 1|5,0.5) $, 
and
$C^{OW}_{0, 0} (\gamma_{gp}, 1|5,0.5) $. Hence, we confirm our conclusion above that the best design choice is the two-way scheme assuming Eve's PNR detection, which also means to engineer the transmission power of the pulse to maximize $C^{TW}_{0, {\rm PNR}} (\gamma_{gp}, 1|\xi,\zeta)$.

\subsection{Choice of wiretap channel model for comparison}
For the comparison between the one-way and two-way regimes,
we need to explain that 
the direction of secure communication is not so important due the following reason.
First, we point out that in a real application, we usually need secure communication in both directions.
Second, the cost of public channel is much smaller than the secure channel
in a usual case\footnote{Here, we briefly explain why the cost of a public channel is much smaller than a secure channel.
A public channel can be realized by conventional wireless channel with error correction,
where we do not care about its secrecy.
When we focus on the secrecy capacity to realize secure communication, 
we need to control the strength power of the pulse to maximize the difference of the mutual information.
However, when using the public channel, we do not need to care about the mutual information to the eavesdropper because we do not care about its secrecy.
Hence, we can make our pulse as strong as possible.
Therefore, the cost of public channel is almost negligible in comparison with the cost of secure communication.}.
In this case, once we can realize secure communication from Alice to Bob,
we can realize secure communication from Bob to Alice
by combining the public channel from Bob to Alice and the secure random seed transmitted from Alice to Bob via the secure communication from Bob to Alice,
which can be realized by use of Shannon's one-time pad secure communication.
Therefore, it is sufficient to establish secure communication with an easier direction. 
That is, it is important to maximize secrecy capacity 
under the assumption that we can freely use the public channel 
that leads to optimal use of available system resources. 

For example, we suppose that 
the wiretap channel from a ground station to a satellite like Fig. \ref{fig:OpticalScenario}
is much better than the wiretap channel from the satellite
to the ground station.
When we employ the one-way protocol,
the secure information transmission from the ground station to the satellite
is much easier than 
that from the satellite to the ground station.
Oppositely, when we employ the two-way protocol,
the secure communication from the satellite to the ground station
is much easier than that from the ground station to the satellite.
In this case, it is natural that we employ only 
the wiretap channel from the ground station to the satellite.
Now, we consider the case that the ground station wants to send a secure message to the satellite.
When we use the one-way scheme, we can simply use a wiretap channel code 
for the wiretap channel from the ground station to the satellite.
When we use the two-way scheme, 
the satellite securely transmits random seeds to the ground station by using our two-way protocol.
Then, the ground station sends the secret message to the satellite
by combining the public channel and the secure random seed
transmitted from the satellite to the ground station.
Hence, in total we need three rounds of communication in this case.
In contrast, 
when we use the wiretap channel from the ground station to the satellite
and the satellite wants to send a secure message to the ground station,
the use of the one-way scheme requires two rounds of communication totally
as follows.
In this case, the ground station transmits random securely seeds to 
the satellite by using the one-way protocol.
Then, the satellite sends the secret message to the ground station
by combining the public channel and the secure random seed
transmitted from the ground station to the satellite.
Therefore, when we employ the wiretap channel from the ground station to the satellite,
it is natural to compare the two-way secrecy capacity from the satellite to the ground station
with the one-way secrecy capacity from the ground station to the satellite.
That is, when we compare the one-way and two-way secrecy capacities,
it is natural to focus on the same Poisson wiretap channel.

\section{Conclusions and further improvements}
In this paper, we have formulated the one-way and two-way secrecy capacities
for Poisson wiretap channel with two types of detectors.
In the first setting,  
the legitimate and the eavesdropper are assumed to have a threshold detector.
In the second setting, 
the eavesdropper is assumed to have a PNR detector.
In this case, 
the legitimate is also assumed to have a PNR detector
in the one-way regime.
However, in the two-way regime, 
the legitimate is assumed to use a threshold detector
because we employ the same protocol in the two-way regime.
At the second round of the two-way regime, 
Alice sends the exclusive or between the encoded information and the received binary variable in the first round.
Due to this construction, we can achieve the secrecy capacity only with linear code over the finite field $\bF_2$, which is an advantage of the two-way regime.
When Bob uses a PNR detector, the secrecy rate can be improved.
However, this case requires more complicated code construction.

Indeed, when the Poisson channel to the legitimate is 
a degraded channel of the Poisson channel to the eavesdropper, 
the one-way secrecy capacities are always zero.
That is, a secure communication is impossible in this case.
Since the above situation is natural in the man-in-the-middle attack,
the above impossibility is a serious defect.
Fortunately, as shown in this paper,
the two-way secrecy capacities are always positive unless
the channel to the eavesdropper is noiseless.
It means that our two-way scheme brings us a secure communication
even when the eavesdropper makes the man-in-the-middle attack
in the communication of the first round. 
Here, the communication of the second round is treated as a public channel.
That is, we assume that the eavesdropper makes the man-in-the-middle attack in the communication of the second round. 
To clarify this advantage, 
we have numerically compared these secrecy capacities.
Our numerical comparison shows drastic improvements by the two-way method 
when the eavesdropper is more powerful than the legitimate.

In order to implement the proposed two-way method in a real communication system, 
we need to perform finite-length analysis for concrete code constructions.
Such a discussion for the Gaussian channel and BPSK modulation has already been performed in the previous paper \cite{arXiv}.
This type of analysis for the proposed method for the Poisson channel is our ongoing study.
Further, our assumption for perfect Poisson channel is ideal.
In practice, the Photoelectric converter and post-processing circuits are typically used, thus the distribution of the received signal would be different from the Poisson distribution \cite{R3,R4,R5}.
An analysis on such a realistic setting is also future study.

In addition, the obtained security does not depend on any assumption
for the calculation power of the computer
because the security analysis is ultimately based on the amount of leaked information to the eavesdropper.
Hence, the proposed method works as post quantum cryptography.
One might consider whether 
our protocol keeps secrecy over 
active man-in-the-middle attacks.
Since active man-in-the-middle attacks are allowed to modify 
the message transmission to the legitimate,
our two-way protocol cannot realize secrecy over such an attack.
Quantum key distribution (QKD) \cite{BB84,Shor} covers secrecy even in this case.
While the original QKD protocol employs the single photon transmission,
the decoy QKD protocol \cite{Hwang,LMC,Wang,H-decoy} offers the secrecy even for the multiple-photon transmission
by using many rounds of communications via the public noiseless channel.

Indeed, one might consider that our two-way protocol is too complicated for practical implementation.
However, the detector to be used in the two-way protocol 
is the threshold detector, which can be the same as that for conventional wireless communication.
That is, 
since the average photon number is not so small,
the required photo detector has low sensitivity (large dark count).
The difference from the conventional one-way protocol is the number of rounds.
On the other hand, the above decoy QKD protocol 
has a protocol that is much more complicated than ours.
Since the average photon number is quite small,
it also requires a special quantum detector, i.e., 
a detector with high sensitivity (small dark count),
which is much more expensive.
Hence, our additional cost is almost negligible in comparison with the additional cost of the above decoy QKD protocol.
Therefore, under the assumption of passive (not active) man-in-the-middle attack, we have shown that our two-way protocol is more efficient in the sense of cost effectiveness and can always be used to complement other security protocols.

\section*{Acknowledgment}
MH was supported in part by JSPS Grant-in-Aid for Scientific Research (A) No.17H01280 and for Scientific Research (B) No.16KT0017, and Kayamori Foundation of Informational Science Advancement.

\appendices
\section{Stochastically degraded condition}\label{A1}
\begin{figure}[tbh]
\centering
\includegraphics[scale=0.37]{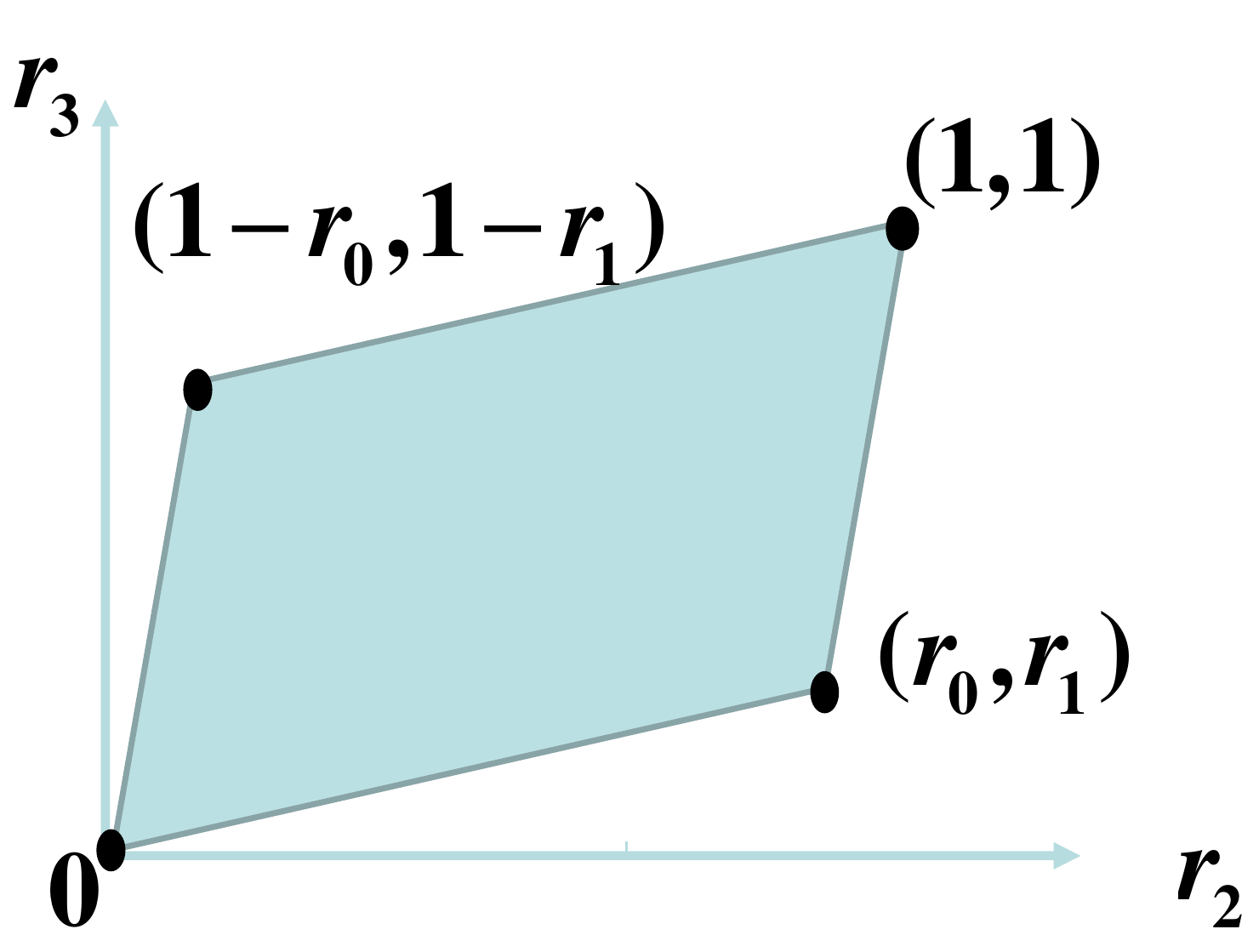}
\protect\caption{
Equivalent condition to stochastically degradedness;
$(r_2,r_3)$ belongs to this tetragon.}
\label{fig:2-dim}
\end{figure}

Now, we consider the condition for 
stochastically degraded property for the one-way case when 
the output systems of Bob and Eve 
are composed a binary data.
To consider this problem, we consider a general binary wiretap channel model
with Alice's binary input $A$, Bob's binary output $B$, and Eve's binary output $E$.
We set the parameters $r_0,r_1,r_2,r_3$ as
$W_{B|A}(0|0)=r_0$, $W_{B|A}(0|1)=r_1$,
$W_{E|A}(0|0)=r_2$, and $W_{E|A}(0|1)=r_3$.
Then, the channel $(W_{B|A},W_{E|A})$ is stochastically degraded
if and only if 
$(r_2,r_3)$ is written 
as $\lambda_1 (r_0, r_1)+\lambda_2(1-r_0, 1-r_1)$
with $\lambda_1,\lambda_2 \in [0,1]$.
This condition is equivalent to the condition that
$(r_2,r_3)$ belongs to the tetragon spanned by
$(0,0)$, $(r_0,r_1)$, 
$(1,1)$, and
$(1-r_1,1-r_1)$ as Fig. \ref{fig:2-dim}.
Assume that $r_0 \ge r_1$ and $r_2 \ge r_3$.
When $ r_2 \le r_0 $, 
the above condition is equivalent to the condition 
$ \frac{r_3}{r_2}\ge \frac{r_1}{r_0}$.
When $ r_2 \ge r_0 $, 
the above condition is equivalent to the condition 
$ \frac{1-r_3}{1-r_2}\le \frac{1-r_1}{1-r_0}$.
In our case, 
\begin{align*}
&r_0=\tau(\zeta,\kappa), \quad r_1=\tau(\xi+ \zeta,\kappa), \\
&r_2=\tau(\gamma_{np}\zeta,\kappa'),\quad
r_3=\tau(\gamma_{gp} \xi+ \gamma_{np}\zeta,\kappa').
\end{align*}
Hence, the conditions $r_0 \ge r_1$ and $r_2 \ge r_3$ always hold.

Now, we consider the case when $\kappa=\kappa'$.
In this case, 
$\tau(a,\kappa)$ is monotonically decreasing for $a$.
Hence, when $\gamma_{np} \ge 1$, we have 
$ r_2 \le r_0 $.
The above condition is equivalent to the condition 
\begin{align}
\frac{\tau(\gamma_{gp} \xi+ \gamma_{np}\zeta,\kappa)}{
\tau(\gamma_{np}\zeta,\kappa)}
\ge
\frac{\tau( \xi+ \zeta,\kappa)}{\tau(\zeta,\kappa)}.\label{Equ1}
\end{align}
Especially, 
when $\gamma_{np}=1$, 
the above condition is simplified to 
\begin{align}
\tau(\gamma_{gp} \xi+ \zeta,\kappa)
\ge
\tau( \xi+ \zeta,\kappa).\label{Equ2}
\end{align}
Since $\tau(a,\kappa)$ is monotonically decreasing for $a$,
it is equivalent to the condition 
$\gamma_{gp}\le 1$.

When $\gamma_{np} \le 1$, we have 
$ r_2 \ge r_0 $.
The above condition is equivalent to the condition 
$ (1-\tau(\gamma_{gp} \xi+ \gamma_{np}\zeta,\kappa))/
 (1-\tau(\gamma_{np}\zeta,\kappa))
\le
 (1-\tau(\xi+ \zeta,\kappa))/
 (1-\tau(\zeta,\kappa))$, i.e.,
\begin{align*}
\begin{aligned}
& \tau(\gamma_{gp} \xi+ \gamma_{np}\zeta,\kappa)
+\tau(\zeta,\kappa)
-\tau(\gamma_{gp} \xi+ \gamma_{np}\zeta,\kappa) \tau(\zeta,\kappa) \\
\ge &
\tau(\gamma_{np}\zeta,\kappa)
+\tau(\xi+ \zeta,\kappa)
-\tau(\xi+ \zeta,\kappa) \tau(\gamma_{np}\zeta,\kappa).
\end{aligned}
\end{align*}

Next, we consider the case when $\kappa=\kappa'=0$,  i.e., the case with the on-off detector.
Since $\tau(a+b,0)=\tau(a,0) \tau(b,0)$,
the condition \eqref{Equ1} under the condition $\gamma_{np} \ge 1$
can be simplified to 
the condition 
$ \tau(\gamma_{gp} \xi,0)\ge \tau( \xi,0)$, i.e.,
$ \gamma_{gp}  \le 1$.

\section{Cardinality of auxiliary random variable}\label{A2}
Here, we show that
when the number $|{\cal X}|$ of elements of ${\cal X}$ is $\ell$,
the maximum in \eqref{eq:secrecy_capacity}
can be achieved with the case of $|{\cal V}|=\ell$.
Assume that $|{\cal V}|>\ell$ and the cardinality of the support of $P_V$
is greater than $\ell$.
Then, there exists a joint distribution $P_{V,U}$ on ${\cal V} \times {\cal U}$
such that
the marginal distribution of $P_{V,U}$ for $V$ equals $P_V$,
the cardinality of the support of $P_{V|U=u}$ is not greater than $\ell$
and $ \sum_{v\in {\cal V}}P_{X|V}(x|v) P_{V|U}(v|u)=  \sum_{v\in {\cal V}}P_{X|V}(x|v) P_{V}(v)$. We have
\begin{align*}
& I(Y;V)-I(Z;V)
=
H(Y)-H(Z)-H(Y|V)+H(Z|V) \\
= & H(Y)-H(Z)
-\sum_{v \in {\cal V}}
P_{V}(v)\big(H(Y|V=v)-H(Z|V=v)\big) \\
= & H(Y)-H(Z)\\
&-
\sum_{u \in {\cal U}}
\sum_{v \in {\cal V}}
P_U(u)P_{V|U}(v|u)\big(H(Y|V=v)-H(Z|V=v)\big) \\
= & 
\sum_{u \in {\cal U}}P_U(u)
\Big(H(Y)-H(Z) \\
& -\sum_{v \in {\cal V}}
P_{V|U}(v|u)\big( H(Y|V=v)-H(Z|V=v)\big) \Big)\\
= & 
\sum_{u \in {\cal U}}P_U(u)
\Big(H(Y|U=u)-H(Z|U=u) \\
&-\big(H(Y|V,U=u)-H(Z|V,U=u)\big) \Big)\\
=&
\sum_{u \in {\cal U}}P_U(u)
\Big(I(Y;V|U=u)-I(Z;V|U=u)\Big).
\end{align*}
Therefore, there exists an element $u \in {\cal U}$
such that
\begin{align*}
 I(Y;V)-I(Z;V)
\le 
I(Y;V|U=u)-I(Z;V|U=u).
\end{align*}
Since 
the cardinality of the support of $P_{V|U=u}$ is not greater than $\ell$,
we obtain the desired statement.

\end{document}